\documentclass{ws-p8-50x6-00}
\newcommand{\unit}[1]{\,\mathrm{#1}}

\begin{document}

\title{Search for $\eta_{\mathrm b}$ in Two-Photon Events}

\author{Armin B\"ohrer}

\address{Fachbereich Physik, Universit\"at Siegen, 57068 Siegen, Germany\\
E-mail: armin.boehrer@cern.ch}


\maketitle

\abstracts{A search for the pseudoscalar meson $\eta_{\mathrm b}$ was 
performed at LEP~II energies with an integrated luminosity of 
$700\unit{pb^{-1}}$. 
The search, done for the decay modes into 4 and 6 charged particles 
yielded 0 and 1 candidates, respectively. Upper limits on 
$\Gamma_{\gamma\gamma}(\eta_{\mathrm b}) \times$BR for both modes of 
$57\unit{eV}$ and $128\unit{eV}$ were obtained with corresponding 
limits of $17\%$ and $38\%$ on branching ratios 
BR($\eta_{\mathrm b} \rightarrow 4$ charged particles) and 
BR($\eta_{\mathrm b} \rightarrow 6$ charged particles) at 
a confidence level of 95\%. The 
candidate has a mass of $9.30 \pm 0.04\unit{GeV}$.}

\section{Introduction and Motivation}

\begin{picture}(10,0)(0,0)
\put(370.,184.){SI-2001-8}
\put(370.,171.){Oct.\ 2001}
\end{picture}

\vspace*{-0.5cm}

The ${{\mathrm b}\bar{\mathrm b}}$ ground state, 
the $\eta_{\mathrm b}$ meson has not been 
observed yet. Because of their initial state two-photon 
collisions are well suited for the study of pseudoscalar mesons 
$J^{PC} = 0^{+-}$, in which they can be produced exclusively. 
The high $\gamma\gamma$ cross section, the high LEP luminosity and 
energy as well as the low background 
from other processes make LEP~II events a good sample to search for 
this meson. 

The ALEPH experiment has started a search for the still undiscovered
$\eta_{\mathrm b}$ pseudoscalar meson\cite{etabconfnote}. The 
meson has been searched for in 
two-photon events in its exclusive decays to 4 and 6 charged particles. 

The analysis is motivated by various predictions for 
the mass of the $\eta_{\mathrm b}$, e.g., from potential models, pQCD,
NRQCD, and lattice calculations, which are tested or constrained by 
a measurement of the $\eta_{\mathrm b}$ mass. See the contributions 
by G.~Bali\cite{bali} and by S.~Collins\cite{collins} at this conference 
and the contributed paper\cite{etabconfnote}. The allowed 
mass $m(\eta_{\mathrm b})$ of these 
estimates ranges from $9.32\unit{GeV}/c^2$ to $9.45\unit{GeV}/c^2$. 

\section{Potential for the Measurement}

The production cross sections has been estimated as follows. 
In two-photon collisions 
the cross section for $\eta_{\mathrm b}$ is calculated with the 
equivalent photon approximation\cite{RPP,budnev} using the estimated 
$\eta_{\mathrm b}$ mass in the form factor (a mass of $9.4\unit{GeV}$ is 
used). The partial width $\Gamma_{\gamma\gamma}$ is calculated from the ratio 
$\Gamma_{\gamma\gamma}(\eta_{\mathrm b})/
\Gamma_{\gamma\gamma}(\eta_{\mathrm c})$ using the Coulomb potential 
approach\cite{yellow96-01}. 

Using the measured partial width 
$\Gamma_{\gamma\gamma}(\eta_{\mathrm c}) = 7.4 \pm 1.4\unit{keV}$ a value of 
$\Gamma_{\gamma\gamma}(\eta_{\mathrm b}) = 416\unit{eV}$ is obtained. 
(See, however, the contribution\cite{fabiano} by N.~Fabiano at this 
conference indicating a larger $\Gamma_{\gamma\gamma}(\eta_{\mathrm b})$.) 
This translates into a production cross section of 
$0.222\unit{pb}$ at $\sqrt{s} = 197
\unit{GeV}$ (luminosity weighted for the $700\unit{pb^{-1}}$ at LEP~II above 
W$^+$W$^-$ threshold). This would correspond to 156 $\eta_{\mathrm b}$ 
mesons produced in ALEPH during LEP~II data taking.

The branching ratios of the $\eta_{\mathrm b}$ decay can only be estimated.
Here, a new (different to the ALEPH conference note\cite{etabconfnote}) 
approach is used, where for the production
probability of $n_{\mathrm{pair}}$ pion (or kaon) pairs a
Poisson distribution is assumed. The number $n_{\mathrm{pair}}$ can be
obtained, because the energy evolution of the mean charged multiplicity
$\langle n \rangle$
is predicted in modified leading log approximation (MLLA) with the
assumption of local parton-hadron duality (LPHD)\cite{lphd} and is
fitted to ${\mathrm e^+}{\mathrm e^-}$ data\cite{alephqcd}.
With the choice $n_{\mathrm{pair}} = \langle n \rangle /2$ for charged
and $n_{\mathrm{pair}} = \langle n \rangle /4$ for neutral pairs,
the branching ratios can be estimated. Evaluation for the $\eta_{\mathrm c}$
a branching ratio to (0 neutral and) 4 charged particles of
$P_0P_2 = 9.9\%$ is estimated, while the
measured decays add up to $9.3 \pm 1.8\%$\cite{RPP}. For the
$\eta_{\mathrm b}$ decay to 4 and 6 charged particles $P_0P_2 = 2.7\%$ and
$P_0P_3 = 3.3\%$ are obtained, respectively.

The selection and reconstruction efficiences are studied using events 
generated with PHOT02\cite{alex}. For the decays it is assumed that 
the momentum distributions are
given by phase space. The efficiencies
are found to be 15.7\% and 10.1\%, respectively.

%\begin{table}[th]
\begin{center}
\begin{tabular}{|lll|}
\hline
mean cms energy      & = & $196.6\unit{GeV}$ \\
total luminosity     & = & $700\unit{pb^{-1}}$\\
$\alpha_s(m(\eta_{\mathrm b}/2))$, $\alpha_s(m(\eta_{\mathrm c}/2))$
                     & = & 0.22, 0.36\\
$\Lambda_{5,4}$      & = & $220\unit{MeV}$, $306\unit{MeV}$\\
cross section        & = & $0.222\unit{pb}$ (lumi.weighted)\\
                     & & \\
\# produced $\eta_{\mathrm b}$ & = & 156\\
                     & & \\
efficiency (4 charged) & = & 15.7\% \\
efficiency (6 charged) & = & 10.1\% \\
                     & & \\
\# $\eta_{\mathrm b}$ (BR(4 charged) = 2.7\%) & = & 0.65 events \\
\# $\eta_{\mathrm b}$ (BR(6 charged) = 3.3\%) & = & 0.52 events \\
\hline
\end{tabular}
%\caption{Numbers important in the analysis.\label{tablenumbers}}
\end{center}
%\end{table}
%%  ~\\

%\vspace{-2cm}

\subsection{Data Analysis}

In order to keep the efficiency high, loose selection cuts are chosen for 
event and track selection. 
So, no attempt is made to reconstruct the K$_{\mathrm S}$ at this stage. The 
mass resolution of the accepted events is around $0.14\unit{GeV}$, mainly due 
to $\pi -$K misidentification and mass assignment, with an 
average reconstructed mass of $9.4\unit{GeV}$. A signal region 
$9.0\unit{GeV}$ to $9.8\unit{GeV}$ is chosen.

The background, which is dominated by 
$\gamma\gamma$ continuum production, is therefore estimated 
from the number of data events: a) the number of events in the signal 
region before the final cuts, which is 78 (139) events for 4 (6) charged 
particles, and b) 
by a fit to the ratio of the mass spectra after all cuts are applied 
and before the final cuts on $\sum p_{t,i}$, thrust, 
$\theta ({\mathrm{thrust~axis}})$, and hemisphere mass are applied. 
The background is estimated to be $0.3 \pm 0.3$ ($0.8 \pm 0.4$) events.

\section{Results}

\begin{figure}
\begin{minipage}{.48\textwidth}
\includegraphics[width=1.00\textwidth,clip]{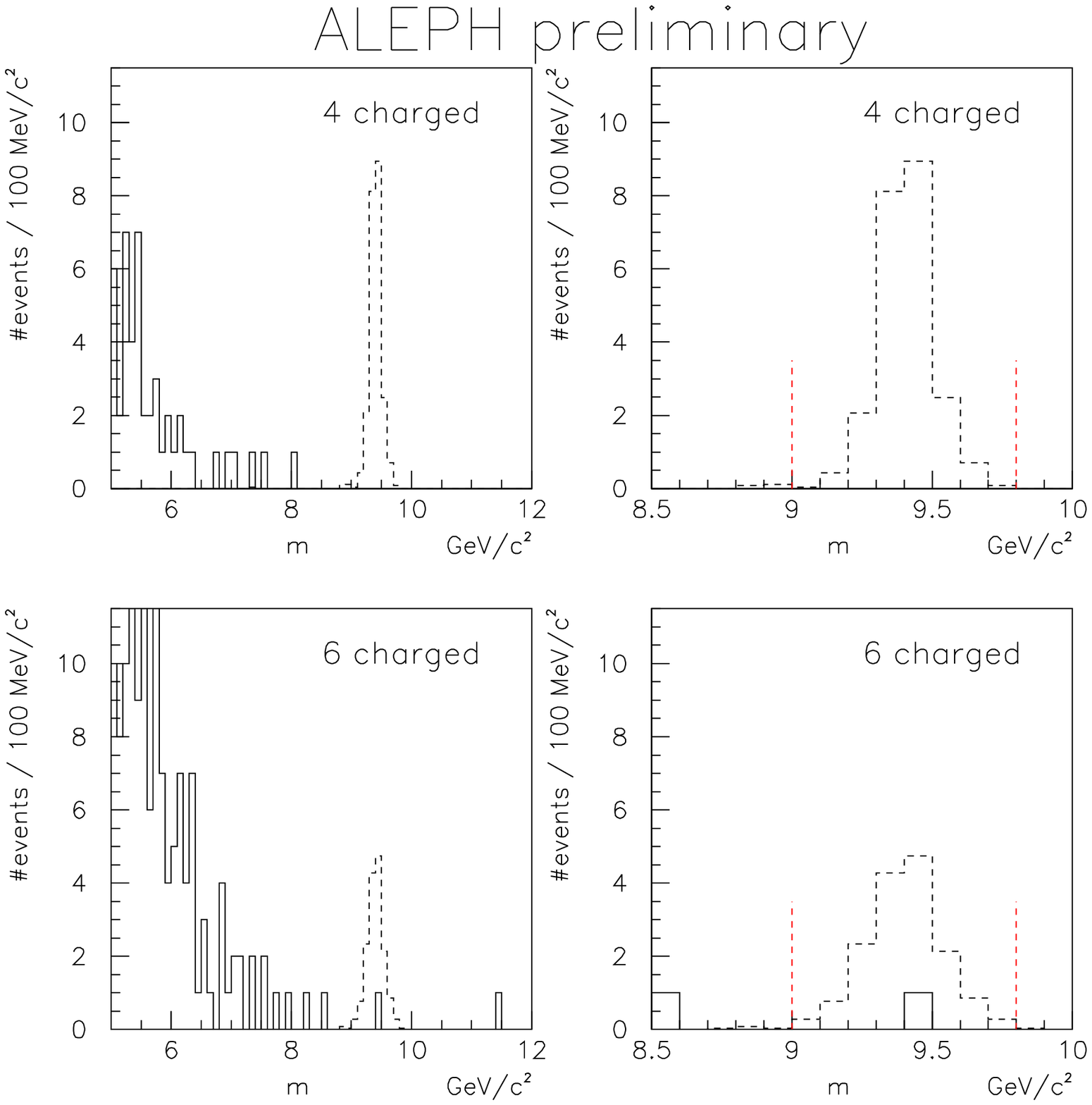}
\caption{Invariant mass distributions in the search for the $\eta_{\mathrm b}$
meson. Data (solid line) are compared to the expected signal for an assumed
branching ratio of 100\% for the decay to 4 and 6 charged particles}
\label{signalmass}
\end{minipage}
\begin{minipage}{.04\textwidth}
~
\end{minipage}
\begin{minipage}{.48\textwidth}
\includegraphics[width=1.00\textwidth,clip]{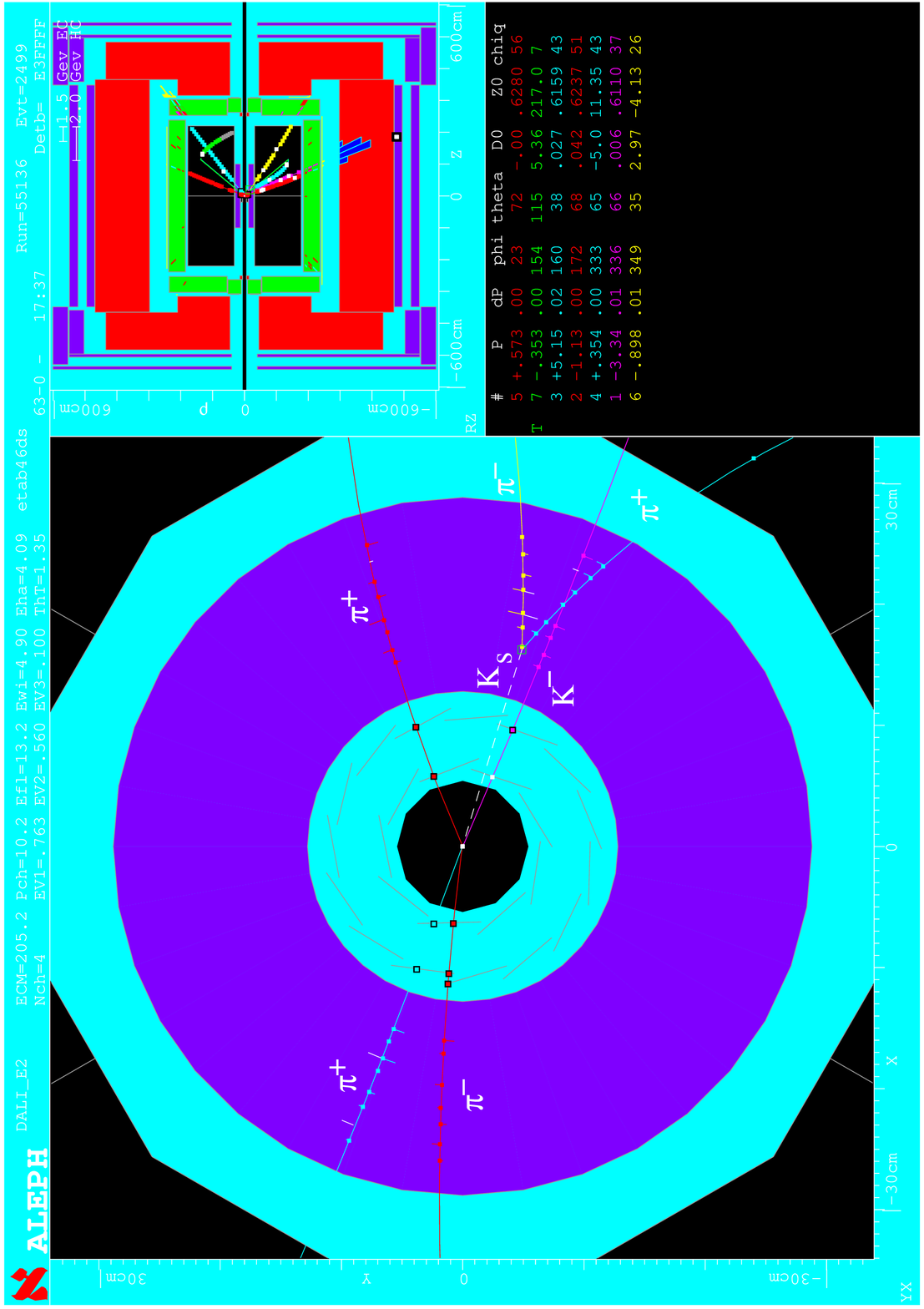}
\caption{An $r\phi$ view of the 
$\eta_{\mathrm b} \rightarrow {\mathrm K}_{\mathrm S}$K$^- \pi^+ \pi^- \pi^+$
candidate event with the reconstructed mass of
$9.30 \pm 0.02 \pm 0.02\unit{GeV}/c^2$,
selected in the signal region}
\label{etabbest-inner}
\end{minipage}
\end{figure}

The invariant mass spectra of the selected events from $700\unit{pb^{-1}}$ at 
LEP~II with 4 (and 6) charged particles are shown in Figure \ref{signalmass}. 
In the signal region 0 (and 1) events are found. 

From the knowledge of the background, the efficiency and 
its uncertainty ($\pm 25 \%$) the observed number 
of events are converted\cite{zech} into upper limits.

For a confidence limit of $\alpha = 95\%$ upper limits  
for $\Gamma_{\gamma\gamma}(\eta_{\mathrm b}) \times$BR of 
$57\unit{eV}$ (and $128\unit{eV}$) are obtained. Including the evaluated 
two photon width of $416\unit{eV}$ and its uncertainty ($\pm 25 \%$) 
the upper limits on the branching ratio of the $\eta_{\mathrm b}$ meson at 
$95\%$ C.L.\ are 
BR($\eta_{\mathrm b} \rightarrow 4$ charged particles) $<$ 17\% and 
BR($\eta_{\mathrm b} \rightarrow 6$ charged particles) $<$ 38\%.

The selected event, shown in Figure \ref{etabbest-inner}, 
has a V$^0$ particle compatible with a K$_{\mathrm S}$. One daughter of the 
V$^0$ has the kaon mass assigned. Compatible in d$E$/d$x$ with a pion 
($\chi^2_{\pi} = 0.86$ rather than $\chi^2_{\mathrm K} = 0.18$) we 
recalculate the mass of the event. A mass estimate of 
$9.30 \pm 0.04\unit{GeV}$ is obtained, where the error is a conservative 
estimate of the total error.

\section{Conclusion}

In an integrated luminosity of $700\unit{pb^{-1}}$ at LEP~II energies 
the pseudoscalar meson $\eta_{\mathrm b}$ has been searched for by 
the ALEPH collaboration in its decays to 4 and 6 charged particles. 
One candidate is retained in the decay mode into 6 charged particles, 
while the expected signal is 0.65 and 0.52 events for the two 
decay modes and a background of $0.3 \pm 0.3$ ($0.8 \pm 0.4$) events is 
expected. The candidate has a mass of
$9.30 \pm 0.04\unit{GeV}$. The event is compatible with background.

Upper limits on 
$\Gamma_{\gamma\gamma}(\eta_{\mathrm b}) \times$BR of 
$57\unit{eV}$ and $128\unit{eV}$ corresponding to limits of the 
branching ratios, 
BR($\eta_{\mathrm b} \rightarrow 4$ charged particles) and 
BR($\eta_{\mathrm b} \rightarrow 6$ charged particles), 
of $17\%$ and $38\%$ are obtained with a confidence level of 95\%. 
A discovery would need the effort of all 4 LEP experiments.

\end{document}